\documentclass[%
 notitlepage,reprint,
 amsmath,amssymb,
 aps,prl
]{revtex4-1}

\usepackage{cases}
\usepackage{graphicx}
\usepackage{dcolumn}
\usepackage{bm}
\usepackage{color}
\usepackage{amsfonts,latexsym}
\usepackage{enumerate}
\usepackage{amsmath}
\usepackage{amssymb}
\usepackage{rotating}
\usepackage[hidelinks]{hyperref}
\usepackage{xcolor}
\hypersetup{
	colorlinks,
	linkcolor={red!50!black},
	citecolor={blue!50!black},
	urlcolor={blue!80!black}
}

\newcommand{\Eq}[1]{Eq.~\eqref{#1}}
\newcommand{\eq}[1]{\eqref{#1}}

\newcommand{\beq}{\begin{equation}}
\newcommand{\eeq}{\end{equation}}

\newcommand{\beqa}{\begin{eqnarray}}
\newcommand{\eeqa}{\end{eqnarray}}

\def\bal#1\eal{\begin{align}#1\end{align}}

\def\Bal#1\Eal{\begin{align*}#1\end{align*}}

\newcommand{\Beqa}{\begin{eqnarray*}}
\newcommand{\Eeqa}{\end{eqnarray*}}

\newcommand{\nn}{\nonumber}

\newcommand{\me}{\mathrm{e}}

\newcommand{\mi}{\mathrm{i}}

\newcommand{\dif}{\mathrm{d}}

\newcommand{\vect}[1]{\mathbf{#1}}

\begin{document}

\title{$d$-dimensional L\"uscher's formula and the near-threshold three-body states in a finite volume}

\author{Shangguo Zhu$^{3}$}
\email{shangguo.zhu@gatech.edu}
\author{Shina Tan$^{1,2,3,4}$}
\email{shinatan@pku.edu.cn}
\affiliation{$^{1}$International Center for Quantum Materials, School of Physics, Peking University, Beijing 100871, China}
\affiliation{$^{2}$Collaborative Innovation Center of Quantum Matter, Beijing 100871, China}
\affiliation{$^{3}$School of Physics, Georgia Institute of Technology, Atlanta, Georgia 30332, USA}
\affiliation{$^{4}$Center for Cold Atom Physics, Chinese Academy of Sciences, Wuhan 430071, China}

\begin{abstract}
	
	We study two particles colliding in a $d$-dimensional finite volume and generalize L\"uscher's formula to arbitrary $d$ spatial dimensions. 
	We obtain the $s$- and $p$-wave approximations of the generalized L\"uscher's formula. 	
	For resonant $s$- or $p$-wave interactions, we analytically determine the energies of the low-lying states at large box size $L$.
	At $s$-wave resonance, we discover two low-lying states with nearly opposite energies, which are proportional to $\pm 1/L^{d/2}$ for $d\ge 5$, or $\pm 1/L^{2}\sqrt{\ln L}$ for $d=4$. 
	This provides important insights into the near-threshold states of three bosons at a three-body resonance in a 2- or higher-dimensional finite volume. 
			
\end{abstract}
\maketitle

\textit{Introduction.}---Many important simulations of the physical systems, such as atomic nuclei, hadrons, and cold atoms, are performed in a finite volume, usually a box with periodic boundary condition. 
The relation between the two-body energy spectrum in the finite volume and the scattering phase shifts is described by L\"uscher's formula~\cite{Luscher1986a,Luscher1986b,Luscher1991}. 
It serves as a standard approach to determining the scattering observables in the finite-volume simulations, including lattice quantum chromodynamics for nuclear systems~\cite{Epelbaum2009}, lattice Monte Carlo for cold atoms~\cite{Endres2011,Endres2013,Lee2008,Lee2008a,Bour2011a} and cold dilute neutron matter~\cite{LeeSchafer}, etc.

The original L\"uscher's formula is specifically derived for two colliding particles in a 3-dimensional (3D) finite volume with cubic symmetry~\cite{Luscher1986a,Luscher1986b,Luscher1991}. 
People have obtained the generalizations of L\"uscher's formula in many different scenarios~\cite{genLuscher}. 
In addition to 3D, L\"uscher's formula has also been derived in 2D~\cite{Fiebig1994} and 1D~\cite{Luscher1990}. 
Beane obtained the $s$-wave approximation of L\"uscher's formula for various $d$~\cite{Beane2010}. 
The formula that Beane obtained is valid if $d \le 3$, but is not convergent if $d\ge 4$ because of the way the lattice sum was regularized.

Although attempts have been made~\cite{Beane2007,Tan2008,Polejaeva2012,Guo2013,Guo2017a,Guo2017,Guo2018,Briceno2013,HansenSharpe,BricenoHansenSharpe,KreuzerHammer,KreuzerRef2,Bour2011,Bour2012,Meissner2015,Meng2018,MaiDoring,Hammer2017two,Doring2018,Romero-Lopez2018,Pang2019}, it is challenging to study a genuine three-body problem in the finite volume and extend L\"uscher's formula to the three-body systems in a succinct manner. 
Kreuzer and Hammer numerically investigated the modification of Efimov spectrum~\cite{Efimov1970,Efimov1970a} below the three-body threshold~\cite{KreuzerHammer}. 
Later, the energy shift of the three-body bound states was analytically calculated~\cite{Meissner2015, Meng2018}. 
Recently, Guo and Gasparian suggested that a multiple-body problem could be mapped to a two-body problem in a higher dimension~\cite{Guo2017}. 
Then, the generalization of L\"{u}scher's formula in higher dimensions would provide important insights into the three-body problems in a 2- or higher-dimensional finite volume.

In this Letter, we study two particles colliding in a $d$D finite volume (a box with side length $L$ and periodic boundary condition) and generalize L\"uscher's formula to arbitrary $d$D in the framework of nonrelativistic quantum mechanics. 
First, we introduce the pseudo wave function, and formulate the problem in the momentum space, where the form of the pseudo wave function remains the same for all spatial dimensions. 
The Fourier transform back to the real space leads to a natural regularization of the lattice sum. 
Then, we obtain the $d$D L\"uscher's formula and its $s$- and $p$-wave approximations. 
The generalized L\"uscher's formula would greatly help the finite-volume simulations of high-dimensional objects in various fields such as string theory.

In addition, for resonant interactions, we analytically calculate the energies of the low-lying states at large box size $L$. 
At an $s$-wave resonance, we discover two low-lying states with nearly opposite energies, whose leading order terms are proportional to $\pm 1/L^{d/2}$ for $d\ge 5$, or $\pm 1/L^{2}\sqrt{\ln L}$ for $d=4$. 
This suggests the existence of two near-threshold states for three bosons at a three-body resonance in a 3D or 2D finite volume, with energies proportional to $ 1/L^3$ or $ 1/L^{2}\sqrt{\ln L}$, respectively. 
The three bosons could be the Efimov trimer~\cite{Efimov1970,Efimov1970a} at the three-body threshold.

\textit{Pseudo wave function.}---We consider two particles with short-range interaction in a $d$D space. 
We assume that their interaction is rotationally invariant and vanishes when their distance $r$ exceeds the range of interaction $r_e^{}$. 
We take the analytically known wave function at $r>r_e^{}$ and its analytic continuation to $r<r_e^{}$ as the \textit{pseudo} wave function: 
\begin{equation}\label{bcreal}
\psi(\vect{r}) = \sum\limits_{l\mu} C_{l \mu}^{}\Big[ j_{dl}^{}(p r) \cot \delta_l(p) - y_{dl}^{}(p r) \Big]Y_{l \mu}^{}(\hat{\vect{r}}).
\end{equation}
Here $p$ is the wave number defined such that the energy $E=\hbar^2 p^2 / m$, with $m$ the mass of each particle and $2\pi\hbar$ the Planck constant.
$l=0,1,2,\cdots$ is the orbital angular momentum. 
$\delta_{l}$ as a function of $p$ is the $l$-wave phase shift in $d$ dimensions~\cite{Hammer2010}. 
$Y_{l \mu}^{}(\hat{\vect{r}})$ denotes the hyperspherical harmonic, normalized as
$\int Y_{l \mu}^{\ast}(\hat{\vect{r}}) Y_{l' \mu'}^{}(\hat{\vect{r}}) \dif^{d-1}\hat{\vect{r}}= \delta_{ll'}^{}\delta_{\mu\mu'}^{}$,
and $\mu$ is a set of quantum numbers $(\mu_1,\mu_2, \ldots, \mu_{d-2})$ belonging to a particular value of $l$ ($d\ge 3$)~\cite{Avery2010}. 
At $d=2$, $Y_{l \mu}^{}(\hat{\vect{r}})$ reduces to the sine and cosine functions. 
$C_{l \mu}^{}$ are some coefficients. 
Also, we have  defined
\begin{equation}
j_{dl}^{}(x)=\sqrt{\frac{\pi}{2}}\frac{J_{l+d/2-1}^{}(x)}{x_{}^{d/2-1}}, \quad
y_{dl}^{}(x)=\sqrt{\frac{\pi}{2}}\frac{Y_{l+d/2-1}^{}(x)}{x_{}^{d/2-1}}, \nn
\end{equation}
where $J_{l}^{}(x)$ and $Y_{l}^{}(x)$ are the Bessel functions of the first and second kind, respectively. 
In the following discussions, we use units such that $m=\hbar=1$.

All the information about the short-range interaction is encapsulated in the phase shifts $\delta_{l}$. 
At low energies, the phase shifts obey the effective range expansion~\cite{Hammer2010}, 
\begin{equation} 
\label{ere_d} 
p^{2l+d-2} \Big[\cot \delta_{l} - \tau_d^{} \frac{2}{\pi}\ln(p R_{dl}^{}) \Big] = -\frac{1}{a_{dl}^{}}+\frac{1}{2}r_{dl}^{}p^2 + O(p^4).
\end{equation} 
Here the parameter $\tau_d^{} = 1$ or $0$ when $d$ is even or odd, respectively. 
$a_{dl}^{}$ is the $l$-wave scattering length with dimension $[{\rm length}]^{2l+d-2}$, 
$r_{dl}^{}$ the $l$-wave effective range with dimension $[{\rm length}]^{-2l-d+4}$, 
and $R_{dl}^{}$ some length scale. 
For $l=0,1,2,\cdots$, ``$l$-wave'' means $s$-wave, $p$-wave, $d$-wave, ..., respectively. 
Then, the short-range interaction is conveniently described by only a few parameters, such as $a_{dl}^{}$, $r_{dl}^{}$, and $R_{dl}^{}$. 
When the scattering length $a_{d,0}^{} \to \pm \infty$ or $a_{d,1}^{} \to \pm \infty$, we say the system is at an $s$- or $p$-wave resonance, respectively.

In the infinite space, $\psi(\vect{r})$ satisfies the Schr\"odinger equation with the actual short-range potential replaced by a pseudo  potential~\cite{Fermi1936,Huang1957,Derevianko2005,Stock2005,Idziaszek2006,Pricoupenko2006}, 
or equivalently the Helmholtz equation with a source of the delta function and its derivatives, 
\begin{equation}\label{helmholtzfs} 
(\nabla^{2}+p^2)\psi(\vect{r})=\sum\limits_{l\mu} \widetilde{C}_{l \mu}^{} Q_{l \mu}^{}(\nabla)\delta(\vect{r}), 
\end{equation} 
where $Q_{l \mu}^{}(\vect{r}) = r^l Y_{l \mu}^{}(\hat{\vect{r}})$ is the $l$-degree harmonic polynomial, and $\widetilde{C}_{l \mu}^{} \equiv C_{l \mu}^{}\frac{(-1)^{l+1}}{\pi}  (2\pi)^{\frac{d+1}{2}} p^{2-d-l}$. 
Due to the source, the solution is singular at the origin. 
Note that any solution plus a smooth solution to the homogeneous equation still solves the above equation. 
The actual solution is uniquely fixed by the ``boundary condition" with the correct phase shifts in \Eq{bcreal}.

Now consider two particles colliding in a $d$D finite volume. 
We focus on their relative motion, assuming that their total momentum is zero.  
The wave function satisfies $\psi(\vect{r})=\psi(\vect{r}+\vect{n}L)$ for all $d$D integral vector $\vect{n}\in \mathbb{Z}^d$. 
Compared to \Eq{helmholtzfs}, the periodic pseudo wave function satisfies the Helmholtz equation with a periodic source
\begin{equation}\label{helmholtzperiodic}
(\nabla^{2}+p^2)\psi(\vect{r})=\sum\limits_{l\mu} \widetilde{C}_{l \mu}^{} Q_{l \mu}^{}(\nabla) \delta_L^{}(\vect{r}) ,
\end{equation}
where $\delta_L^{}(\vect{r}) \equiv \sum_{\vect{n}\in \mathbb{Z}^d}\delta(\vect{r}-\vect{n}L)$ is the periodic delta function.

It is convenient to formulate this problem in the momentum space. 
We find the solution of \Eq{helmholtzperiodic} in the form of a regularized Fourier transform
\begin{align}
\psi(\vect{r})&=\lim\limits_{\epsilon\to 0^{+}} \int\frac{\dif^{d} k}{(2\pi)^d}~ \widetilde{\psi}(\vect{k})\me^{\mi \vect{k}\cdot\vect{r}-\epsilon k^2}, \label{psir}\\
\widetilde{\psi}(\vect{k})&= \sum\limits_{l\mu} \widetilde{C}_{l \mu}^{} Q_{l\mu}^{}(\mi \vect{k}) \frac{ I(\vect{k})}{-k^2+p^2}, \label{psik}
\end{align}
where $I(\vect{k})=\sum_{\vect{n}\in \mathbb{Z}^d}(2\pi/L)^d \delta(\vect{k}-2\pi\vect{n}/L)$ is the Fourier transform of $\delta_L^{}(\vect{r})$. 
The regularization factor $\me^{-\epsilon k^2}$ makes the integral well-defined and naturally provides a regularization of the lattice sum after integration. 
Here we have assumed that the wave number does not take the singular values, namely, $p\ne 2\pi n/L$ for any $\vect{n}\in \mathbb{Z}^d$. 
The case of the singular values can be addressed using similar techniques as in Ref.~\cite{Luscher1991}.

\textit{Quantization condition.}---If $r<L$, the pseudo wave function $\psi(\vect{r})$ has a convergent partial wave expansion as in \Eq{bcreal}.
We project $\psi(\vect{r})$ to the partial wave channels, and by equating it with the partial wave expansion in \Eq{bcreal}, we find 
\begin{align}\label{speceq1}
\sum\limits_{l'\mu'} \Big\{ & \frac{2~ \mi^{l-l'} }{\pi q^{l+l'+d-2}} m_{l\mu,l'\mu'}^{} \nn\\
&- \delta_{ll'}^{}\delta_{\mu\mu'}^{}\Big[\cot \delta_{l}-\mi  \theta(-p^2)\Big]\Big\}C_{l'\mu'}^{}=0,
\end{align}
where
\begin{align}\label{matrixm}
m_{l\mu,l'\mu'}^{}= \lim_{\epsilon\to 0^{+}}\Big[ & \sum_{\vect{n}\in \mathbb{Z}^d}\frac{Q_{l\mu}^{\ast}(\vect{n})Q_{l'\mu'}^{}(\vect{n})}{n^2-q^2}\me^{-\epsilon (n^2-q^2) } \nn\\
&-{\rm P}\int \dif^{d}n~ \frac{Q_{l\mu}^{\ast}(\vect{n})Q_{l'\mu'}^{}(\vect{n})}{n^2-q^2}\me^{-\epsilon (n^2-q^2) }\Big].
\end{align} 
The symbol ${\rm P}$ means taking the principal value~\cite{principavalue}, the symbol $^{\ast}$ denotes the complex conjugate, and the dimensionless parameter $q \equiv p L /2 \pi$.
$\theta(x)$ is the Heaviside step function. 
$\theta(x) = 1$ if $x>0$, and $\theta(x) = 0$ if $x<0$. 
When the energy $p^2<0$, we assume $p=\mi \kappa$ with $\kappa>0$. 
Here we have included the factor $\exp(\epsilon q^2) $ in the right hand side of \Eq{matrixm} so that it converges exponentially fast with error $O[\exp (- \pi^2 / \epsilon ) ]$, according to the Poisson summation formula.

Let $\ell \equiv (l\mu)$ be a collective index. 
In order for \Eq{speceq1} to have a nontrivial solution for the coefficients $\{ C_{l'\mu'}^{} \}$, we obtain the following quantization condition: 
\begin{equation}\label{spectrum}
\det \mathcal{M}=0, 
\end{equation}
where $\mathcal{M}$ is an infinite-dimensional matrix with matrix elements 
\begin{equation}\label{matm}
\mathcal{M}_{\ell\ell'}^{} = \frac{2~ \mi^{l-l'} }{\pi q^{l+l'+d-2}} m_{\ell\ell'}^{}-\delta_{\ell\ell'}^{}\Big[\cot \delta_{\ell}-\mi  \theta(-p^2)\Big],
\end{equation}
$\ell'\equiv (l'\mu')$, $\delta_{\ell\ell'}^{} \equiv \delta_{ll'}^{}\delta_{\mu\mu'}^{}$, and $\delta_{\ell} \equiv \delta_{l}$. 
Equations~\eq{spectrum} and \eq{matm} are the generalization of L\"uscher's formula~\cite{Luscher1991} to arbitrary spatial dimensions.

\textit{$s$-wave approximation.}---We consider two particles colliding in the finite volume. 
The lowest possible partial wave channel is the $s$-wave ($l=0$). 
We neglect all the other channels, assuming $C_{l\mu}^{}=0$ for $l \ge 1$ in \Eq{speceq1}.  
Then, $\mathcal{M}$ simply reduces to a scalar, and the quantization condition becomes 
\begin{equation} 
\label{swavespectrum}
\frac{ \pi^{d/2+1} }{ \Gamma(d/2) } q^{d-2} \Big[ \cot \delta_{0}(p)-\mi  \theta(-p^2) \Big] = S_{d}^{}(q^2),
\end{equation} 
where $\Gamma(x)$ is the gamma function and 
\begin{equation}\label{sdx}
S_{d}^{}(x)=\lim_{\epsilon\to 0^{+}}\Big[ \sum_{\vect{n}\in \mathbb{Z}^d}\frac{\me^{-\epsilon (n^2-x)}}{n^2-x}-{\rm P}\int \dif^{d}n~ \frac{\me^{-\epsilon (n^2-x)}}{n^2-x}\Big].
\end{equation} 
Here we have used $Q_{0,0}^{}(\vect{n}) = \sqrt{\frac{\Gamma(d/2)}{2\pi^{d/2}}}$. 
A similar formula with a different regularization was derived by Beane~\cite{Beane2010}, but it is not convergent if $d\ge 4$~\cite{Beane2010}. 
If $d\le 3$, \Eq{swavespectrum} agrees with Beane's formula and other previous results~\cite{Luscher1986a,Luscher1991,Fiebig1994,Beane2004}.

\begin{figure}[t!]
	\includegraphics[scale=0.29]{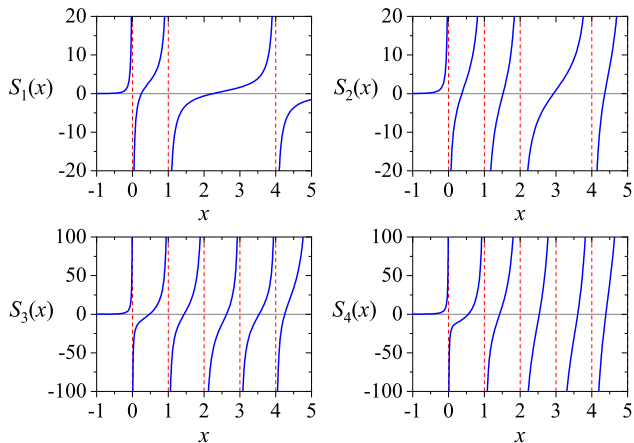}
	\caption{\label{fig_sdz} 
		The function $S_{d}^{}(x)$ for $d=1$, $2$, $3$, and $4$. 
		It exponentially decays when $x$ is large and negative.
		The red dashed lines indicate the singularities (simple poles at $x>0$), where $x$ is equal to the norm square of a integral vector in $\mathbb{Z}^d$. 
	}
\end{figure}

In Fig.~\ref{fig_sdz} we plot the functions $S_d^{}(x)$ for $d\le4$.
$S_{d}^{}(x)$ is singular when $x$ is equal to the norm square of an integral vector in $\mathbb{Z}^d$. 
We see from \Eq{sdx} that the singularities are simple poles when $x>0$. 
For $d=1$, we find an analytical expression $S_{1}^{}(x)=-\frac{\pi}{\sqrt{x}}\cot(\pi \sqrt{x}) -\frac{\pi}{\sqrt{-x}}\theta(-x)$.

When $x$ is large and negative, $S_{d}^{}(x)$ is exponentially small and has the following asymptotic expansion 
\begin{equation} 
S_{d}^{}(-\xi^2)=2d\pi \xi^{\frac{d-3}{2}}\me^{-2\pi \xi}+O(\me^{-2\sqrt{2}\pi \xi}),
\end{equation} 
where $\xi\to +\infty$. 
If the two particles have a bound state in infinite space, 
when they are placed in the large finite volume, $q^2$ corresponding to the bound state is large and negative, and is of order $L^2$. 
The correction to the bound state energy is exponentially small, of order $ L^{(1-d)/2}\me^{-|p| L }$. 
This small correction was previously derived for the 3D two-body systems~\cite{Luscher1986a,Beane2004}. 
For three-body systems, the exponentially small correction to the binding energy in the finite volume was also studied~\cite{Meissner2015,Meng2018}.

The energy eigenvalues of the noninteracting particles in a finite volume are simply $E_{\vect{n}} = (2\pi n /L )^2 $ with $\vect{n}\in \mathbb{Z}^d$. 
For off-resonant interactions, the energy eigenvalues are slightly modified by higher-order terms in powers of $1/L$. 
For the low-lying state with $n=0$, the leading order term is $E = \frac{u_{d}^{} a_{d,0}^{}}{2 L^d}$ for $d\ge 3$, and $E = \frac{2\pi}{L^2\ln (L/2\pi \widetilde{a}_{2,0}^{}) }$ for $d = 2$. 
Here $u_d^{} = 2^{d+1}\pi^{d/2-1}\Gamma(d/2)$ and $\widetilde{a}_{2,0}^{} \equiv  R_{2,0}^{} \exp(-\pi/2a_{2,0}^{})$ is the reduced scattering length in 2D.

It is particularly interesting to investigate the system with resonant interactions. 
We focus on the $s$-wave resonance ($a_{d,0}^{} \to \pm\infty$) in $d\ge4$ dimensions. 
In the infinite space, there is an $s$-wave bound state with vanishing energy. 
The corresponding wave function behaves like $1/r^{d-2}$ at $r>r_e^{}$. 
Naively, one might expect that the correction to the energy due to the finite volume is proportional to $1/L^{d-2}$, since the periodic boundary condition mainly affects the wave function at $r\sim L$. 
However, quite counterintuitively, we discovered \emph{two} low-lying states with nearly opposite energies. 
We derived their energies at large $L$ using \Eq{swavespectrum}:\\ 
\begin{subequations}\label{Eres}\indent 1) for $d=4$,
\begin{align} \label{Eresd4}
E_{\pm}^{} =& \pm\frac{2\sqrt{2}\pi}{L^2 \ln^{1/2}(L/ 2\pi \widetilde{r}_{4,0}^{})} -\frac{\alpha_{4,1}^{}}{L^2 \ln(L/ 2\pi \widetilde{r}_{4,0}^{})} \nn\\
&+ O\Big( \frac{1}{L^2\ln^{3/2}L}\Big),
\end{align} 
\indent 2) for $d\ge5$,
\begin{equation} \label{Eresd5}
E_{\pm}^{} = \pm\frac{ u_{d}^{1/2}}{(-r_{d,0}^{})^{1/2}}\frac{1}{L^{d/2}} + \frac{u_{d}^{}\alpha_{d,1}^{}}{8\pi^2 r_{d,0}^{}}\frac{1}{L^{d-2}} + \cdots, 
\end{equation}
\end{subequations} 
where $\widetilde{r}_{4,0}^{} \equiv  R_{4,0}^{}\exp(\pi r_{4,0}^{}/4)$ is the reduced effective range. 
Note that $r_{d,0}^{}<0$ if $d \ge 5$~\cite{Hammer2010}. 
The parameter $\alpha_{d,1}^{}$ (with $d>2$) is defined as a regularized lattice sum
\begin{equation}
\alpha_{d,1}^{}=\lim_{\epsilon\to 0^{+}}\Big(\sum_{\vect{n}\in \mathbb{Z}^d, \vect{n}\ne \vect{0}}\frac{\me^{-\epsilon n^2}}{n^{2}}\Big) - \epsilon - \frac{2\pi^{d/2}}{d-2}\epsilon^{1-d/2} .
\end{equation}
At small $\epsilon$, this formula has error $O(\me^{-\pi^2/\epsilon})$ and therefore is convenient for numerical evaluation~\cite{dos}. 
We list two values $\alpha_{4,1}^{}\approx -5.54518$ and $\alpha_{6,1}^{}\approx -3.37968$.

To understand why we have two low-lying states, we perform a perturbative analysis of the pseudo wave function in two regions: 
1) when $\vect{r}$ is fixed and $L\to \infty$, 
and 2) when $\vect{r}$ and $L$ both go to infinity with the ratio $\vect{r}/L$ fixed. 
In Region 1), we solve \Eq{helmholtzperiodic} (with $\widetilde{C}_{l\mu}=0$ for $l\ge 1$) perturbatively at small energies and for resonant $s$-wave interaction, and find if $d=4$, 
\begin{subequations}
\label{psismallp}
\begin{equation}
    \psi = \frac{1}{r^2} + \frac{1}{2} p^2 \ln \frac{2  \me^{1/2-\gamma_{\rm E}^{}} \widetilde{r}_{4,0}^{}}{ r }  + O(p^4),
\end{equation}
where $\gamma_{\rm E}^{} \approx 0.577216 $ is the Euler constant, and if $d\ge 5$,
\begin{align}
    \psi =& \frac{1}{r^{d-2}} + p^2 \Big[ \frac{1}{2(d-4)r^{d-4}} \nn\\
    &+ \frac{\pi r_{d,0}^{} }{2^{d-1}\Gamma(d/2-1)\Gamma(d/2)}\Big]+ O(p^4).
\end{align}
\end{subequations}
In Region 2), we solve \Eq{helmholtzperiodic} (with $\widetilde{C}_{l\mu}=0$ for $l\ge 1$) to find
\begin{equation}
\label{trialpsi}
    \psi = \sum\limits_{\vect{n}\in \mathbb{Z}^d} \frac{\kappa^{d-2}}{2^{d/2-2}\Gamma(d/2-1) } \frac{K_{d/2-1}(\kappa r_{\vect{n}}^{})}{(\kappa r_{\vect{n}}^{})^{d/2-1}},
\end{equation}
where $K_{l}(x)$ is the modified Bessel function of the second kind and $\vect{r}_{\vect{n}}^{} \equiv \vect{r}-\vect{n}L $. 
This solution works for both $p^2<0$ and $p^2>0$.
When $p^2>0$, $\kappa = - \mi p $ with $p > 0 $. 
Assuming $|\kappa L| \ll 1$ and using the tail-singularity separation technique~\cite{Tan2008}, we can calculate the discrete sum in \Eq{trialpsi} approximately to find that
if $d=4$,
\begin{subequations}
\label{trialpsiexpansions}
\begin{align}
    \psi =& \sideset{}{'}\sum_{\vect{n}} \Big( \frac{1}{r_{\vect{n}}^{2} } +\frac{1}{2}\kappa^2 \ln r_{\vect{n}}^{}  \Big) \me^{-\epsilon  r_{\vect{n}}^{2}} \nn\\
    &+ \frac{4\pi^2 }{\kappa^2 L^4} + O(\kappa^4 L^2),
\end{align}
and if $d\ge 5$,
\begin{align}
    \psi =&  \sideset{}{'}\sum_{\vect{n}} \Big( \frac{1}{r_{\vect{n}}^{d-2} } - \frac{\kappa^2 }{ 2(d-4)r_{\vect{n}}^{d-4}} \Big) \me^{-\epsilon  r_{\vect{n}}^{2}}
    \nn\\
    &+ \frac{4\pi^{d/2} }{\Gamma(d/2-1)\kappa^2 L^d} + O(\kappa^4 L^{6-d}),
\end{align}
\end{subequations}
where $\epsilon \to 0^{+}$ and $\sideset{}{'}\sum\limits_{\vect{n}} \equiv \sum\limits_{\vect{n}\in \mathbb{Z}^d} - \int \dif^d n $. 
Then, by matching the two expansions of the wave function in the overlap region $r_e^{}\ll r \ll L$, we obtain an equation for the energy: if $d=4$, 
\begin{subequations}
\label{approxenergy}
\begin{equation}
    \frac{4\pi^2}{\kappa^2 L^4} \approx -\frac{1}{2} \kappa^2 \ln \frac{\widetilde{r}_{4,0}^{}}{L}, 
\end{equation}
and if $d\ge 5$,
\begin{equation}
    \frac{4\pi^{d/2} }{\Gamma(d/2-1)\kappa^2 L^d} \approx - \kappa^2 \frac{\pi r_{d,0}^{}}{2^{d-1}\Gamma(d/2-1)\Gamma(d/2)}. 
\end{equation}
\end{subequations}
Solving \Eq{approxenergy}, we indeed find two energies and they are consistent with the leading order terms in \Eq{Eres}.

The three-body problem in $w$ dimensions in the center-of-mass frame can be mapped to a one-body problem in $2w$ dimensions. The interaction between three particles can be expressed as $V_{2}^{}(r_{12}^{}) + V_{2}^{}(r_{23}^{}) + V_{2}^{}(r_{31}^{})+ V_{3}^{}(r_{12}^{}, r_{23}^{}, r_{31}^{})$, where $\vect{r}_{ij}^{} = \vect{r}_{i}^{}-\vect{r}_{j}^{}$ is the position vector between particles $i$ and $j$, and $V_{2}^{}$ and $V_{3}^{}$ are the two-body and three-body potentials, respectively.  
$V_{3}^{}$ has a short range and vanishes when any of the three distances exceeds a certain range. 
Then, the problem is mapped to 6D (4D) by defining the vector $\vect{r} = \big(\vect{r}_{23}^{},(\vect{r}_{12}^{} - \vect{r}_{31}^{})/\sqrt{3}  \big)$ if the three particles live in 3D (2D). 
In $2w$ dimensions, $V_2^{}$ is mapped to a long-range potential, but $V_{3}^{}$ is still mapped to a short-range potential.

For three-body systems with nonzero $V_{3}^{}$ and vanishing $V_{2}^{}$, the $s$-wave scattering length $a_{6,0}^{}$ or $a_{4,0}^{}$ is equivalent to a numerical constant times the three-body scattering hypervolume, the parameter characterizing the effective three-body interaction~\cite{Tan2008,Zhu2017}. 
The $s$-wave resonance in 6D ($a_{6,0}^{} \to \pm\infty$) or 4D ($a_{4,0}^{} \to \pm\infty$) exactly corresponds to the three-body resonance in 3D or 2D, respectively. 
Then, the results in \Eq{Eres} are applicable to these three-body systems.

The systems with vanishing $V_{2}^{}$ can be experimentally approximated in cold atoms. 
The two-body $s$-wave scattering length can be tuned to a zero crossing through Feshbach resonance~\cite{Chin2010,zerocrossing}, and the two-body interaction can then be neglected. 
However, the three-body interaction is usually off-resonance at the same time.  
New mechanisms of tuning the three-body interaction independently are needed to study the standalone three-body resonance effects.

For most other cases, the two-body interaction is present at the three-body resonance.   
For example, when the two-body scattering length is much larger than the range of interaction and is tuned to some critical negative value, an Efimov trimer reaches the three-body resonance, with vanishing binding energy~\cite{Braaten2007}.

For three-body systems at a three-body resonance with a finite two-body scattering length, the three-body physics becomes dominant. 
The results in \Eq{Eres} provide a good qualitative description. 
They suggest that two states exist near the threshold with energies proportional to $ 1/L^3$ or $ 1/L^2 \sqrt{\ln L}$, for three bosons at a three-body resonance in a 3D or 2D finite volume, respectively. 
The proportionality constants depend on the two-body scattering length and a parameter (the counterpart of $r_{6,0}^{}$ or $\widetilde{r}_{4,0}^{}$) from the three-body physics~\cite{TanPrivateNote}.

\textit{$p$-wave approximation.}---According to Eqs.~\eq{speceq1} and \eq{matrixm}, the even-$l$ and odd-$l$ partial waves are decoupled. 
Now we focus on the sector of odd $l$'s, namely $l=1,3,5,\dots$.
If we neglect the scattering phase shifts for $l=3,5,\dots$, we only need to consider the $p$-wave scattering.
Using the formula $Q_{1,\mu}^{}(\vect{r})=[\pi^{-d/2}\Gamma(d/2+1)]^{1/2}r_{\mu}^{}$, where $\mu=1,2,\cdots,d$ and $r_{\mu}^{}$ is the $\mu$th Cartesian component of $\vect{r}$, 
we simplify \Eq{spectrum} as
\begin{equation}\label{pwavespectrum1}
\frac{ \pi^{d/2+1} }{ \Gamma(d/2)}q^{d-2}  \Big[ \cot \delta_{1}(p)-\mi  \theta(-p^2) \Big] = S_{d}^{}(q^2).
\end{equation}
Each solution for $p^2$ corresponds to a $d$-fold degenerate level.

At a $p$-wave resonance ($a_{d,1}^{}\to \pm\infty$), we find one $d$-fold degenerate low-lying level with energy $E$ as follows.\\
\begin{subequations}\label{Epwaveres}\indent 1) If $d=2$,
	\begin{align} \label{Epwaveresd2}
	E = &\frac{2\pi}{L^2 \ln(L/ 2\pi \widetilde{r}_{2,1}^{})} - \frac{\alpha_{2,1}^{}}{L^2 \ln^{2}(L/ 2\pi \widetilde{r}_{2,1}^{})} \nn\\
	&+ O\Big( \frac{1}{L^2\ln^{3}L}\Big) .
	\end{align} 
	\indent 2) If $d\ge3$,
	\begin{equation} \label{Epwaveresdg3}
	E = \frac{u_{d}^{}}{-r_{d,1}^{}L^d} - \frac{u_{d}^{2}\alpha_{d,1}^{}}{4\pi^2  r_{d,1}^{2} L^{2d-2}} + \cdots.
	\end{equation}
\end{subequations} 
Here $\alpha_{2,1}^{}= \lim\limits_{\Lambda\to +\infty}\Big(\sum\limits_{\vect{n}\in \mathbb{Z}^{2}, \vect{n}\ne \vect{0}}^{n<\Lambda} \frac{1}{n^2}\Big) - 2\pi \ln \Lambda = \lim\limits_{\epsilon\to 0^+}\Big(\sum\limits_{\vect{n}\in \mathbb{Z}^{2}, \vect{n}\ne \vect{0}} \frac{\me^{-\epsilon n^2}}{ n^{2} }\Big) - \epsilon + \pi \ln \epsilon + \pi \gamma_{\rm E}^{}  \approx 2.58498 $ 
and $\widetilde{r}_{2,1}^{} \equiv  R_{2,1}^{}\exp(\pi r_{2,1}^{}/4)$ is the reduced $p$-wave effective range. 
Note that $r_{d,1}^{} <0 $ if $d\ge 3$~\cite{Hammer2010}.

\textit{Summary.}---We studied two particles colliding in a finite volume (a box with periodic boundary condition) and generalized L\"uscher's formula to arbitrary $d$ dimensions. 
We derived the $s$- and $p$-wave approximations of the quantization condition for the energy. 
The corrections to the energies of bound states in the finite volume are exponentially small.  
At $s$-wave resonance, we found two low-lying states for $d \ge 4$.  
The energies of the two states are proportional to $\pm 1/L^{d/2}$ for $d\ge 5$, or $\pm 1/L^{2}\sqrt{\ln L}$ for $d=4$. 
This implies that there exist two near-threshold states with energies proportional to $1/L^{3}$ or $1/L^{2}\sqrt{\ln L}$, for three bosons at a three-body resonance in a 3D or 2D finite volume, respectively.

The volume dependence of the energies can be readily checked and verified by numerical simulations, such as diffusion quantum Monte Carlo. 
The generalized L\"uscher's formula and the analytical results about the low-lying states would greatly help the study of high-dimensional systems and three-body systems (such as the Efimov trimers) in finite volume.

\begin{acknowledgments}
	We gratefully acknowledge support by the National Science Foundation CAREER Award Grant No. PHY-1352208. This research was supported in part by the National Science Foundation under Grant No. NSF PHY-1125915. 
\end{acknowledgments}


\begin{thebibliography}{53}
	\expandafter\ifx\csname natexlab\endcsname\relax\def\natexlab#1{#1}\fi
	\expandafter\ifx\csname bibnamefont\endcsname\relax
	\def\bibnamefont#1{#1}\fi
	\expandafter\ifx\csname bibfnamefont\endcsname\relax
	\def\bibfnamefont#1{#1}\fi
	\expandafter\ifx\csname citenamefont\endcsname\relax
	\def\citenamefont#1{#1}\fi
	\expandafter\ifx\csname url\endcsname\relax
	\def\url#1{\texttt{#1}}\fi
	\expandafter\ifx\csname urlprefix\endcsname\relax\def\urlprefix{URL }\fi
	\providecommand{\bibinfo}[2]{#2}
	\providecommand{\eprint}[2][]{\url{#2}}
	
	\bibitem[{\citenamefont{L\"{u}scher}(1986{\natexlab{a}})}]{Luscher1986a}
	\bibinfo{author}{\bibfnamefont{M.}~\bibnamefont{L\"{u}scher}},
	\bibinfo{journal}{Commun. Math. Phys.} \textbf{\bibinfo{volume}{104}},
	\bibinfo{pages}{177} (\bibinfo{year}{1986}{\natexlab{a}}).
	
	\bibitem[{\citenamefont{L\"{u}scher}(1986{\natexlab{b}})}]{Luscher1986b}
	\bibinfo{author}{\bibfnamefont{M.}~\bibnamefont{L\"{u}scher}},
	\bibinfo{journal}{Commun. Math. Phys.} \textbf{\bibinfo{volume}{105}},
	\bibinfo{pages}{153} (\bibinfo{year}{1986}{\natexlab{b}}).
	
	\bibitem[{\citenamefont{L{\"{u}}scher}(1991)}]{Luscher1991}
	\bibinfo{author}{\bibfnamefont{M.}~\bibnamefont{L{\"{u}}scher}},
	\bibinfo{journal}{Nucl. Phys. B} \textbf{\bibinfo{volume}{354}},
	\bibinfo{pages}{531} (\bibinfo{year}{1991}).
	
	\bibitem[{\citenamefont{Epelbaum et~al.}(2009)\citenamefont{Epelbaum, Hammer,
			and Mei\ss{}ner}}]{Epelbaum2009}
	\bibinfo{author}{\bibfnamefont{E.}~\bibnamefont{Epelbaum}},
	\bibinfo{author}{\bibfnamefont{H.-W.} \bibnamefont{Hammer}},
	\bibnamefont{and} \bibinfo{author}{\bibfnamefont{U.-G.}
		\bibnamefont{Mei\ss{}ner}}, \bibinfo{journal}{Rev. Mod. Phys.}
	\textbf{\bibinfo{volume}{81}}, \bibinfo{pages}{1773} (\bibinfo{year}{2009}).
	
	\bibitem[{\citenamefont{Endres et~al.}(2011)\citenamefont{Endres, Kaplan, Lee,
			and Nicholson}}]{Endres2011}
	\bibinfo{author}{\bibfnamefont{M.~G.} \bibnamefont{Endres}},
	\bibinfo{author}{\bibfnamefont{D.~B.} \bibnamefont{Kaplan}},
	\bibinfo{author}{\bibfnamefont{J.-W.} \bibnamefont{Lee}}, \bibnamefont{and}
	\bibinfo{author}{\bibfnamefont{A.~N.} \bibnamefont{Nicholson}},
	\bibinfo{journal}{Phys. Rev. A} \textbf{\bibinfo{volume}{84}},
	\bibinfo{pages}{043644} (\bibinfo{year}{2011}).
	
	\bibitem[{\citenamefont{Endres et~al.}(2013)\citenamefont{Endres, Kaplan, Lee,
			and Nicholson}}]{Endres2013}
	\bibinfo{author}{\bibfnamefont{M.~G.} \bibnamefont{Endres}},
	\bibinfo{author}{\bibfnamefont{D.~B.} \bibnamefont{Kaplan}},
	\bibinfo{author}{\bibfnamefont{J.-W.} \bibnamefont{Lee}}, \bibnamefont{and}
	\bibinfo{author}{\bibfnamefont{A.~N.} \bibnamefont{Nicholson}},
	\bibinfo{journal}{Phys. Rev. A} \textbf{\bibinfo{volume}{87}},
	\bibinfo{pages}{023615} (\bibinfo{year}{2013}).
	
	\bibitem[{\citenamefont{Lee}(2008{\natexlab{a}})}]{Lee2008}
	\bibinfo{author}{\bibfnamefont{D.}~\bibnamefont{Lee}}, \bibinfo{journal}{Phys.
		Rev. C} \textbf{\bibinfo{volume}{78}}, \bibinfo{pages}{024001}
	(\bibinfo{year}{2008}{\natexlab{a}}).
	
	\bibitem[{\citenamefont{Lee}(2008{\natexlab{b}})}]{Lee2008a}
	\bibinfo{author}{\bibfnamefont{D.}~\bibnamefont{Lee}}, \bibinfo{journal}{Eur.
		Phys. J. A} \textbf{\bibinfo{volume}{35}}, \bibinfo{pages}{171}
	(\bibinfo{year}{2008}{\natexlab{b}}).
	
	\bibitem[{\citenamefont{Bour et~al.}(2011{\natexlab{a}})\citenamefont{Bour, Li,
			Lee, Mei\ss{}ner, and Mitas}}]{Bour2011a}
	\bibinfo{author}{\bibfnamefont{S.}~\bibnamefont{Bour}},
	\bibinfo{author}{\bibfnamefont{X.}~\bibnamefont{Li}},
	\bibinfo{author}{\bibfnamefont{D.}~\bibnamefont{Lee}},
	\bibinfo{author}{\bibfnamefont{U.-G.} \bibnamefont{Mei\ss{}ner}},
	\bibnamefont{and} \bibinfo{author}{\bibfnamefont{L.}~\bibnamefont{Mitas}},
	\bibinfo{journal}{Phys. Rev. A} \textbf{\bibinfo{volume}{83}},
	\bibinfo{pages}{063619} (\bibinfo{year}{2011}{\natexlab{a}}).
	
	\bibitem[{Lee()}]{LeeSchafer}
	\bibinfo{note}{D. Lee and T. Sch\"afer, Phys. Rev. C \textbf{73}, 015201
		(2006); \textbf{73}, 015202 (2006)}.
	
	\bibitem[{gen()}]{genLuscher}
	\bibinfo{note}{X. Li and C. Liu, Phys. Lett. B \textbf{587}, 100 (2004); X.
		Feng, X. Li, and C. Liu, Phys. Rev. D \textbf{70}, 014505 (2004); K.
		Rummukainen and S. Gottlieb, Nucl. Phys. B \textbf{450}, 397 (1995); C. Kim,
		C. Sachrajda, and S. R. Sharpe, Nucl. Phys. B \textbf{727}, 218 (2005); M.
		G\"ockeler \textit{et al.}, Phys. Rev. D \textbf{86}, 094513 (2012); C. Liu,
		X. Feng, and S. He, Int. J. Mod. Phys. A \textbf{21}, 847 (2006); V. Bernard
		\textit{et al.}, J. High Energy Phys. 01 (2011) 019 ; N. Li and C. Liu, Phys.
		Rev. D \textbf{87}, 014502 (2013); P. F. Bedaque and J. W. Chen, Phys. Lett.
		B \textbf{616}, 208 (2005); A. Agadjanov \textit{et al.}, Nucl. Phys.
		\textbf{B886}, 1199 (2014).}
	
	\bibitem[{\citenamefont{Fiebig et~al.}(1994)\citenamefont{Fiebig, Woloshyn, and
			Dominguez}}]{Fiebig1994}
	\bibinfo{author}{\bibfnamefont{H.}~\bibnamefont{Fiebig}},
	\bibinfo{author}{\bibfnamefont{R.}~\bibnamefont{Woloshyn}}, \bibnamefont{and}
	\bibinfo{author}{\bibfnamefont{A.}~\bibnamefont{Dominguez}},
	\bibinfo{journal}{Nucl. Phys. B} \textbf{\bibinfo{volume}{418}},
	\bibinfo{pages}{649} (\bibinfo{year}{1994}).
	
	\bibitem[{\citenamefont{L{\"{u}}scher and Wolff}(1990)}]{Luscher1990}
	\bibinfo{author}{\bibfnamefont{M.}~\bibnamefont{L{\"{u}}scher}}
	\bibnamefont{and} \bibinfo{author}{\bibfnamefont{U.}~\bibnamefont{Wolff}},
	\bibinfo{journal}{Nucl. Phys. B} \textbf{\bibinfo{volume}{339}},
	\bibinfo{pages}{222} (\bibinfo{year}{1990}).
	
	\bibitem[{\citenamefont{Beane}(2010)}]{Beane2010}
	\bibinfo{author}{\bibfnamefont{S.~R.} \bibnamefont{Beane}},
	\bibinfo{journal}{Phys. Rev. A} \textbf{\bibinfo{volume}{82}},
	\bibinfo{pages}{063610} (\bibinfo{year}{2010}).
	
	\bibitem[{\citenamefont{Beane et~al.}(2007)\citenamefont{Beane, Detmold, and
			Savage}}]{Beane2007}
	\bibinfo{author}{\bibfnamefont{S.~R.} \bibnamefont{Beane}},
	\bibinfo{author}{\bibfnamefont{W.}~\bibnamefont{Detmold}}, \bibnamefont{and}
	\bibinfo{author}{\bibfnamefont{M.~J.} \bibnamefont{Savage}},
	\bibinfo{journal}{Phys. Rev. D} \textbf{\bibinfo{volume}{76}},
	\bibinfo{pages}{074507} (\bibinfo{year}{2007}).
	
	\bibitem[{\citenamefont{Tan}(2008)}]{Tan2008}
	\bibinfo{author}{\bibfnamefont{S.}~\bibnamefont{Tan}}, \bibinfo{journal}{Phys.
		Rev. A} \textbf{\bibinfo{volume}{78}}, \bibinfo{pages}{013636}
	(\bibinfo{year}{2008}).
	
	\bibitem[{\citenamefont{Polejaeva and Rusetsky}(2012)}]{Polejaeva2012}
	\bibinfo{author}{\bibfnamefont{K.}~\bibnamefont{Polejaeva}} \bibnamefont{and}
	\bibinfo{author}{\bibfnamefont{A.}~\bibnamefont{Rusetsky}},
	\bibinfo{journal}{Eur. Phys. J. A} \textbf{\bibinfo{volume}{48}},
	\bibinfo{pages}{67} (\bibinfo{year}{2012}).
	
	\bibitem[{\citenamefont{Guo}()}]{Guo2013}
	\bibinfo{author}{\bibfnamefont{P.}~\bibnamefont{Guo}},
	\eprint{arXiv:1303.3349}.
	
	\bibitem[{\citenamefont{Guo}(2017)}]{Guo2017a}
	\bibinfo{author}{\bibfnamefont{P.}~\bibnamefont{Guo}}, \bibinfo{journal}{Phys.
		Rev. D} \textbf{\bibinfo{volume}{95}}, \bibinfo{pages}{054508}
	(\bibinfo{year}{2017}).
	
	\bibitem[{\citenamefont{Guo and Gasparian}(2017)}]{Guo2017}
	\bibinfo{author}{\bibfnamefont{P.}~\bibnamefont{Guo}} \bibnamefont{and}
	\bibinfo{author}{\bibfnamefont{V.}~\bibnamefont{Gasparian}},
	\bibinfo{journal}{Phys. Lett. B} \textbf{\bibinfo{volume}{774}},
	\bibinfo{pages}{441} (\bibinfo{year}{2017}).
	
	\bibitem[{\citenamefont{Guo and Gasparian}(2018)}]{Guo2018}
	\bibinfo{author}{\bibfnamefont{P.}~\bibnamefont{Guo}} \bibnamefont{and}
	\bibinfo{author}{\bibfnamefont{V.}~\bibnamefont{Gasparian}},
	\bibinfo{journal}{Phys. Rev. D} \textbf{\bibinfo{volume}{97}},
	\bibinfo{pages}{014504} (\bibinfo{year}{2018}).
	
	\bibitem[{\citenamefont{Brice\~no and Davoudi}(2013)}]{Briceno2013}
	\bibinfo{author}{\bibfnamefont{R.~A.} \bibnamefont{Brice\~no}}
	\bibnamefont{and} \bibinfo{author}{\bibfnamefont{Z.}~\bibnamefont{Davoudi}},
	\bibinfo{journal}{Phys. Rev. D} \textbf{\bibinfo{volume}{88}},
	\bibinfo{pages}{094507} (\bibinfo{year}{2013}).
	
	\bibitem[{Han()}]{HansenSharpe}
	\bibinfo{note}{M. T. Hansen and S. R. Sharpe, Phys. Rev. D \textbf{90}, 116003
		(2014); \textbf{92}, 114509 (2015); \textbf{93}, 014506 (2016); \textbf{93},
		096006 (2016); \textbf{95}, 034501 (2017); S. R. Sharpe, Phys. Rev. D
		\textbf{96}, 054515 (2017); arXiv:1901.00483}.
	
	\bibitem[{Bri()}]{BricenoHansenSharpe}
	\bibinfo{note}{R. A. Brice\~no, M. T. Hansen, and S. R. Sharpe, Phys. Rev. D
		\textbf{95}, 074510 (2017); \textbf{98}, 014506 (2018); \textbf{99}, 014516
		(2019).}
	
	\bibitem[{Kre({\natexlab{a}})}]{KreuzerHammer}
	\bibinfo{note}{S. Kreuzer and H.-W. Hammer, Phys. Lett. B \textbf{673}, 260
		(2009); Eur. Phys. J. A \textbf{43}, 229 (2010)}.
	
	\bibitem[{Kre({\natexlab{b}})}]{KreuzerRef2}
	\bibinfo{note}{S. Kreuzer and H.-W. Hammer, Phys. Lett. B \textbf{694}, 424
		(2011); S. Kreuzer and H. W. Grie\ss{}hammer, Eur. Phys. J. A \textbf{48}, 93
		(2012)}.
	
	\bibitem[{\citenamefont{Bour et~al.}(2011{\natexlab{b}})\citenamefont{Bour,
			K\"onig, Lee, Hammer, and Mei\ss{}ner}}]{Bour2011}
	\bibinfo{author}{\bibfnamefont{S.}~\bibnamefont{Bour}},
	\bibinfo{author}{\bibfnamefont{S.}~\bibnamefont{K\"onig}},
	\bibinfo{author}{\bibfnamefont{D.}~\bibnamefont{Lee}},
	\bibinfo{author}{\bibfnamefont{H.-W.} \bibnamefont{Hammer}},
	\bibnamefont{and} \bibinfo{author}{\bibfnamefont{U.-G.}
		\bibnamefont{Mei\ss{}ner}}, \bibinfo{journal}{Phys. Rev. D}
	\textbf{\bibinfo{volume}{84}}, \bibinfo{pages}{091503}
	(\bibinfo{year}{2011}{\natexlab{b}}).
	
	\bibitem[{\citenamefont{Bour et~al.}(2012)\citenamefont{Bour, Hammer, Lee, and
			Mei\ss{}ner}}]{Bour2012}
	\bibinfo{author}{\bibfnamefont{S.}~\bibnamefont{Bour}},
	\bibinfo{author}{\bibfnamefont{H.-W.} \bibnamefont{Hammer}},
	\bibinfo{author}{\bibfnamefont{D.}~\bibnamefont{Lee}}, \bibnamefont{and}
	\bibinfo{author}{\bibfnamefont{U.-G.} \bibnamefont{Mei\ss{}ner}},
	\bibinfo{journal}{Phys. Rev. C} \textbf{\bibinfo{volume}{86}},
	\bibinfo{pages}{034003} (\bibinfo{year}{2012}).
	
	\bibitem[{\citenamefont{Mei\ss{}ner et~al.}(2015)\citenamefont{Mei\ss{}ner,
			R\'{i}os, and Rusetsky}}]{Meissner2015}
	\bibinfo{author}{\bibfnamefont{U.-G.} \bibnamefont{Mei\ss{}ner}},
	\bibinfo{author}{\bibfnamefont{G.}~\bibnamefont{R\'{i}os}}, \bibnamefont{and}
	\bibinfo{author}{\bibfnamefont{A.}~\bibnamefont{Rusetsky}},
	\bibinfo{journal}{Phys. Rev. Lett.} \textbf{\bibinfo{volume}{114}},
	\bibinfo{pages}{091602} (\bibinfo{year}{2015}).
	
	\bibitem[{\citenamefont{Meng et~al.}(2018)\citenamefont{Meng, Liu, Mei\ss{}ner,
			and Rusetsky}}]{Meng2018}
	\bibinfo{author}{\bibfnamefont{Y.}~\bibnamefont{Meng}},
	\bibinfo{author}{\bibfnamefont{C.}~\bibnamefont{Liu}},
	\bibinfo{author}{\bibfnamefont{U.-G.} \bibnamefont{Mei\ss{}ner}},
	\bibnamefont{and} \bibinfo{author}{\bibfnamefont{A.}~\bibnamefont{Rusetsky}},
	\bibinfo{journal}{Phys. Rev. D} \textbf{\bibinfo{volume}{98}},
	\bibinfo{pages}{014508} (\bibinfo{year}{2018}).
	
	\bibitem[{Mai()}]{MaiDoring}
	\bibinfo{note}{M. Mai and M. D\"oring, Eur. Phys. J. A \textbf{53}, 240 (2017);
		Phys. Rev. Lett. \textbf{122}, 062503 (2019).}
	
	\bibitem[{Ham()}]{Hammer2017two}
	\bibinfo{note}{H.-W. Hammer, J.-Y. Pang, and A. Rusetsky, J. High Energy Phys.
		09 (2017) 109; J. High Energy Phys. 10 (2017) 115.}
	
	\bibitem[{\citenamefont{D\"oring et~al.}(2018)\citenamefont{D\"oring, Hammer,
			Mai, Pang, Rusetsky, and Wu}}]{Doring2018}
	\bibinfo{author}{\bibfnamefont{M.}~\bibnamefont{D\"oring}},
	\bibinfo{author}{\bibfnamefont{H.-W.} \bibnamefont{Hammer}},
	\bibinfo{author}{\bibfnamefont{M.}~\bibnamefont{Mai}},
	\bibinfo{author}{\bibfnamefont{J.-Y.} \bibnamefont{Pang}},
	\bibinfo{author}{\bibfnamefont{A.}~\bibnamefont{Rusetsky}}, \bibnamefont{and}
	\bibinfo{author}{\bibfnamefont{J.}~\bibnamefont{Wu}}, \bibinfo{journal}{Phys.
		Rev. D} \textbf{\bibinfo{volume}{97}}, \bibinfo{pages}{114508}
	(\bibinfo{year}{2018}).
	
	\bibitem[{\citenamefont{Romero-L{\'o}pez
			et~al.}(2018)\citenamefont{Romero-L{\'o}pez, Rusetsky, and
			Urbach}}]{Romero-Lopez2018}
	\bibinfo{author}{\bibfnamefont{F.}~\bibnamefont{Romero-L{\'o}pez}},
	\bibinfo{author}{\bibfnamefont{A.}~\bibnamefont{Rusetsky}}, \bibnamefont{and}
	\bibinfo{author}{\bibfnamefont{C.}~\bibnamefont{Urbach}},
	\bibinfo{journal}{Eur. Phys. J. C} \textbf{\bibinfo{volume}{78}},
	\bibinfo{pages}{846} (\bibinfo{year}{2018}).
	
	\bibitem[{\citenamefont{Pang et~al.}()\citenamefont{Pang, Wu, Hammer, Meißner,
			and Rusetsky}}]{Pang2019}
	\bibinfo{author}{\bibfnamefont{J.-Y.} \bibnamefont{Pang}},
	\bibinfo{author}{\bibfnamefont{J.-J.} \bibnamefont{Wu}},
	\bibinfo{author}{\bibfnamefont{H.-W.} \bibnamefont{Hammer}},
	\bibinfo{author}{\bibfnamefont{U.-G.} \bibnamefont{Meißner}},
	\bibnamefont{and} \bibinfo{author}{\bibfnamefont{A.}~\bibnamefont{Rusetsky}},
	\eprint{arXiv:1902.01111}.
	
	\bibitem[{\citenamefont{Efimov}(1970{\natexlab{a}})}]{Efimov1970}
	\bibinfo{author}{\bibfnamefont{V.~N.} \bibnamefont{Efimov}},
	\bibinfo{journal}{Yad. Fiz.} \textbf{\bibinfo{volume}{12}},
	\bibinfo{pages}{1080} (\bibinfo{year}{1970}{\natexlab{a}}).
	
	\bibitem[{\citenamefont{Efimov}(1970{\natexlab{b}})}]{Efimov1970a}
	\bibinfo{author}{\bibfnamefont{V.}~\bibnamefont{Efimov}},
	\bibinfo{journal}{Phys. Lett. B} \textbf{\bibinfo{volume}{33}},
	\bibinfo{pages}{563} (\bibinfo{year}{1970}{\natexlab{b}}).
	
	\bibitem[{\citenamefont{Hammer and Lee}(2010)}]{Hammer2010}
	\bibinfo{author}{\bibfnamefont{H.-W.} \bibnamefont{Hammer}} \bibnamefont{and}
	\bibinfo{author}{\bibfnamefont{D.}~\bibnamefont{Lee}}, \bibinfo{journal}{Ann.
		Phys. (N. Y.)} \textbf{\bibinfo{volume}{325}}, \bibinfo{pages}{2212}
	(\bibinfo{year}{2010}).
	
	\bibitem[{\citenamefont{Avery}(2010)}]{Avery2010}
	\bibinfo{author}{\bibfnamefont{J.~S.} \bibnamefont{Avery}},
	\bibinfo{journal}{J. Comput. Appl. Math.} \textbf{\bibinfo{volume}{233}},
	\bibinfo{pages}{1366} (\bibinfo{year}{2010}).
	
	\bibitem[{Fer()}]{Fermi1936}
	\bibinfo{note}{E. Fermi, Ric. Sci. {\bf 7-II}, 13 (1936)}.
	
	\bibitem[{\citenamefont{Huang and Yang}(1957)}]{Huang1957}
	\bibinfo{author}{\bibfnamefont{K.}~\bibnamefont{Huang}} \bibnamefont{and}
	\bibinfo{author}{\bibfnamefont{C.~N.} \bibnamefont{Yang}},
	\bibinfo{journal}{Phys. Rev.} \textbf{\bibinfo{volume}{105}},
	\bibinfo{pages}{767} (\bibinfo{year}{1957}).
	
	\bibitem[{\citenamefont{Derevianko}(2005)}]{Derevianko2005}
	\bibinfo{author}{\bibfnamefont{A.}~\bibnamefont{Derevianko}},
	\bibinfo{journal}{Phys. Rev. A} \textbf{\bibinfo{volume}{72}},
	\bibinfo{pages}{044701} (\bibinfo{year}{2005}).
	
	\bibitem[{\citenamefont{Stock et~al.}(2005)\citenamefont{Stock, Silberfarb,
			Bolda, and Deutsch}}]{Stock2005}
	\bibinfo{author}{\bibfnamefont{R.}~\bibnamefont{Stock}},
	\bibinfo{author}{\bibfnamefont{A.}~\bibnamefont{Silberfarb}},
	\bibinfo{author}{\bibfnamefont{E.~L.} \bibnamefont{Bolda}}, \bibnamefont{and}
	\bibinfo{author}{\bibfnamefont{I.~H.} \bibnamefont{Deutsch}},
	\bibinfo{journal}{Phys. Rev. Lett.} \textbf{\bibinfo{volume}{94}},
	\bibinfo{pages}{023202} (\bibinfo{year}{2005}).
	
	\bibitem[{\citenamefont{Idziaszek and Calarco}(2006)}]{Idziaszek2006}
	\bibinfo{author}{\bibfnamefont{Z.}~\bibnamefont{Idziaszek}} \bibnamefont{and}
	\bibinfo{author}{\bibfnamefont{T.}~\bibnamefont{Calarco}},
	\bibinfo{journal}{Phys. Rev. Lett.} \textbf{\bibinfo{volume}{96}},
	\bibinfo{pages}{013201} (\bibinfo{year}{2006}).
	
	\bibitem[{\citenamefont{Pricoupenko}(2006)}]{Pricoupenko2006}
	\bibinfo{author}{\bibfnamefont{L.}~\bibnamefont{Pricoupenko}},
	\bibinfo{journal}{Phys. Rev. A} \textbf{\bibinfo{volume}{73}},
	\bibinfo{pages}{012701} (\bibinfo{year}{2006}).
	
	\bibitem[{pri()}]{principavalue}
	\bibinfo{note}{Alternatively, taking the principal value can be equivalently
		understood as changing the divergent factor $\frac{1}{-k^2+p^2} \to
		\frac{1}{2} \Big(\frac{1}{-k^2+p^2 + \mi \varepsilon} + \frac{1}{-k^2+p^2 -
			\mi \varepsilon} \Big) $, with $\varepsilon$ a positive infinitesimal.}
	
	\bibitem[{\citenamefont{Beane et~al.}(2004)\citenamefont{Beane, Bedaque,
			Parre{\~{n}}o, and Savage}}]{Beane2004}
	\bibinfo{author}{\bibfnamefont{S.~R.} \bibnamefont{Beane}},
	\bibinfo{author}{\bibfnamefont{P.}~\bibnamefont{Bedaque}},
	\bibinfo{author}{\bibfnamefont{A.}~\bibnamefont{Parre{\~{n}}o}},
	\bibnamefont{and} \bibinfo{author}{\bibfnamefont{M.~J.}
		\bibnamefont{Savage}}, \bibinfo{journal}{Phys. Lett. B}
	\textbf{\bibinfo{volume}{585}}, \bibinfo{pages}{106} (\bibinfo{year}{2004}).
	
	\bibitem[{dos()}]{dos}
	\bibinfo{note}{If $d$ is large, the numerical evaluation can be boosted with
		the aid of the ``density of states" function $f(N,d)$, which counts the
		number of $d$-dimensional integral vectors whose length is $n=\sqrt{N}$,
		where $N$ is a non-negative integer. $f(N,d)$ can be obtained quickly using a
		recursive method starting from $d=1$.}
	
	\bibitem[{\citenamefont{Zhu and Tan}()}]{Zhu2017}
	\bibinfo{author}{\bibfnamefont{S.}~\bibnamefont{Zhu}} \bibnamefont{and}
	\bibinfo{author}{\bibfnamefont{S.}~\bibnamefont{Tan}},
	\eprint{arXiv:1710.04147}.
	
	\bibitem[{\citenamefont{Chin et~al.}(2010)\citenamefont{Chin, Grimm, Julienne,
			and Tiesinga}}]{Chin2010}
	\bibinfo{author}{\bibfnamefont{C.}~\bibnamefont{Chin}},
	\bibinfo{author}{\bibfnamefont{R.}~\bibnamefont{Grimm}},
	\bibinfo{author}{\bibfnamefont{P.}~\bibnamefont{Julienne}}, \bibnamefont{and}
	\bibinfo{author}{\bibfnamefont{E.}~\bibnamefont{Tiesinga}},
	\bibinfo{journal}{Rev. Mod. Phys.} \textbf{\bibinfo{volume}{82}},
	\bibinfo{pages}{1225} (\bibinfo{year}{2010}).
	
	\bibitem[{zer()}]{zerocrossing}
	\bibinfo{note}{S. L. Cornish \textit{et al.}, Phys. Rev. Lett. \textbf{85},
		1795 (2000); L. Khaykovich \textit{et al.}, Science \textbf{296}, 1290
		(2002); K. E. Strecker \textit{et al.}, Nature (London) \textbf{417}, 150
		(2002); G. Roati \textit{et al.}, Phys. Rev. Lett. \textbf{99}, 010403
		(2007); T. Lahaye \textit{et al.}, Nature (London) \textbf{448}, 672 (2007);
		M. Fattori \textit{et al.}, Phys. Rev. Lett. \textbf{100}, 080405 (2008); S.
		E. Pollack \textit{et al.}, Phys. Rev. Lett. \textbf{102}, 090402 (2009); Z.
		Shotan \textit{et al.}, Phys. Rev. Lett. \textbf{113}, 053202 (2014).}
	
	\bibitem[{\citenamefont{Braaten and Hammer}(2007)}]{Braaten2007}
	\bibinfo{author}{\bibfnamefont{E.}~\bibnamefont{Braaten}} \bibnamefont{and}
	\bibinfo{author}{\bibfnamefont{H.-W.} \bibnamefont{Hammer}},
	\bibinfo{journal}{Ann. Phys. (N. Y.)} \textbf{\bibinfo{volume}{322}},
	\bibinfo{pages}{120} (\bibinfo{year}{2007}).
	
	\bibitem[{Tan()}]{TanPrivateNote}
	\bibinfo{note}{For three bosons with nonzero two-body interactions at the
		three-body resonance in a 3D finite volume, the leading order formulas of the
		energies ($\propto 1/L^3$) of the two near-threshold states have been derived
		by Shina Tan in his private note.}
	
\end{thebibliography}
\end{document}